\documentstyle[aps,prd,times,psfig]{revtex}
\begin{document}
\draft

\title{Properties of $r$~modes in rotating magnetic
	neutron stars.\\ 
	I. Kinematic Secular Effects and Magnetic
	Evolution Equations.}

\author{Luciano~Rezzolla}
\address{SISSA, International School for Advanced Studies,
         Trieste, Italy.}
\address{INFN, Department of Physics, University of Trieste, Italy.}

\author{Frederick L. Lamb$^{\rm a,b,c}$,  
    	Dragoljub Markovi\'c$^{\rm a,b}$, and 
	Stuart L. Shapiro$^{\rm a,b,c,d}$}

\address{$^{\rm a}$Center for Theoretical Astrophysics,
         University of Illinois at Urbana-Champaign,
         Urbana, Illinois.}
\address{$^{\rm b}$Department of Physics, University of
         Illinois at Urbana-Champaign, Urbana, Illinois.}
\address{$^{\rm c}$Department of Astronomy, University of
         Illinois at Urbana-Champaign, Urbana, Illinois.}
\address{$^{\rm d}$NCSA, University of Illinois at
         Urbana-Champaign, Urbana, Illinois.}

\medskip

\date{\today}
\maketitle

\begin{abstract}
\noindent The instability of \hbox{$r$-mode} oscillations
in rapidly rotating neutron stars has attracted attention
as a potential mechanism for producing high frequency,
almost periodic gravitational waves. The analyses carried
so far have shown the existence of these modes and have
considered damping by shear and bulk viscosity. However,
the magnetohydrodynamic coupling of the modes with a
stellar magnetic field and its role in the damping of the
instability has not been fully investigated
yet. Following our introductory paper~\cite{rls00}, we
here discuss in more detail the existence of secular
higher-order kinematical effects which will produce
toroidal fluid drifts. We also define the sets of
equations that account for the time evolution of the
magnetic fields produced by these secular velocity fields
and show that the magnetic fields produced can reach
equipartition in less than a year. The full numerical
calculations as well as the evaluation of the impact of
strong magnetic fields on the onset and evolution of the
\hbox{$r$-mode} instability will be presented in a
companion paper~\cite{rlms01b}.
\end{abstract}



\section{Introduction}
\label{intro}

	The basic properties of \hbox{$r$-mode}
oscillations in Newtonian rotating stars were
investigated in 1978 by Papaloizou and
Pringle~\cite{pp78} in an attempt to explain the
short-period oscillations seen in cataclysmic variables
in terms of non-radial oscillations of rotating white
dwarfs. Subsequent work~\cite{pbr81,s82} has increased
our understanding of these oscillations, which have close
similarities with the Rossby waves observed in the
Earth's atmosphere and oceans. Almost twenty years later,
\hbox{$r$-mode} oscillations have become the focus of
renewed attention when they were shown to be unstable to
the emission of gravitational radiation. The first
calculations in this sense were carried out by
Andersson~\cite{a98} and by Friedman and
Morsink~\cite{fm98}. Since then, the literature on the
subject has been growing rapidly. Exhaustive reviews of
our present understanding and of unresolved issues
concerning the \hbox{$r$-mode} instability can be found
in~\cite{ak00} and in~\cite{fl00}.

	A significant difference from previously
investigated mode-instabilities is that gravitational
radiation couples with \hbox{$r$-mode} oscillations
primarily through time-varying mass-current multipole
moments rather than through the usual time-varying mass
multipole moments. Coherent, large-scale fluid currents
in a hot plasma with high electrical conductivity, such
as neutron star matter, represent the basic conditions
for the generation of large scale magnetic fields. This
paper is devoted to studying whether intense magnetic
fields can be produced as a result of the onset and
growth of the \hbox{$r$-mode} instability. In particular,
we here extend the discussion first presented
in~\cite{rls00} about the existence of secular velocity
fields which interact with the magnetic field
pre-existing in the neutron star. We show in more detail
that such secular effects can be derived with confidence
from linearized expressions and are responsible for
differential rotation, both in the radial and in the
polar directions. As a result, secular toroidal drifts
appear on isobaric surfaces and generate large-scale
magnetic fields. We also derive here the sets of
equations necessary for the calculation of the magnetic
fields produced in this way. In a companion
paper~\cite{rlms01b} (hereafter paper II) we will present
in detail the results of the numerical solution of the
equations presented here and comment on the impact that
the large magnetic fields produced will have on the onset
and development of the instability.

	The organization of the paper is as follows:
Section~\ref{pp} introduces the physical framework of our
approach and synthesizes the results.
Section~\ref{kinematics}, takes a closer look at the
linear equations of motion for fiducial fluid elements on
isobaric surfaces. Starting from those we then derive the
higher-order secular velocity field and comment on the
validity of our results. Section~\ref{evol_eqs} is
devoted to the formulation of the set of equations
necessary for the calculation of the magnetic field
produced by the secular velocity field derived in the
previous Section. In particular, we discuss two different
approaches and comment on the corresponding advantages
and shortcomings. Section~\ref{conclusions} finally
presents our conclusions and refers the reader to the
numerical results and their astrophysical consequences
which are presented in paper II.

\section{Physical Picture}
\label{pp}

	The energy budget governing the evolution of the
\hbox{$r$-mode} instability is traditionally assumed to
be regulated only by the emission of gravitational waves
and viscosity, which act as sources and sinks of energy,
respectively~\cite{lom98,aks99,oetal98} (In a frame
corotating with the star, the emission of gravitational
waves is seen as an input of energy for the
mode.). However, it is natural to expect that other
sources and sinks should intervene in this budget, most
notably the loss of rotational and mode energy to
electromagnetic radiation~\cite{hl99} and to the coupling
of the mass-currents produced by the oscillations with
the magnetic field present in the neutron
star~\cite{rls00}. In the case of the $r$-mode
instability, this is particularly relevant. The reason
for this is twofold: firstly, because the oscillations
produce large scale mass-currents and, secondly, because
the generation of magnetic field is a generic feature of
shearing flows perpendicular to magnetic field lines in a
highly conducting plasma, such as hot neutron star
matter. This reflects the fact that, in such conditions,
the magnetic field is predominantly advected with the
fluid. When the electrical conductivity is infinite (the
ideal MHD limit), magnetic field lines are ``frozen'' in
the fluid and move entirely with it (Alfv\`en
Theorem,~\cite{p79}).  As pointed out by
Spruit~\cite{s98}, the generation of magnetic fields
under these circumstances can be so efficient that
extremely intense magnetic fields could be created during
the instability. When these become buoyant-unstable they
can then generate powerful flashes of $\gamma$-rays.

	This paper, similarly to the one preceding
it~\cite{rls00} and the one following~\cite{rlms01b},
aims at determining the strength of this coupling and the
consequences that it will have on the onset and evolution
of the instability. In particular, we will show that the
kinematic properties of the \hbox{$r$-mode} oscillations
will give rise to a secular velocity field which, once
coupled to any seed magnetic field, will produce
exponentially growing magnetic fields as the instability
develops.  A detailed analysis of this coupling, which
inevitably introduces a change in the character of the
modes, is beyond the scope of the present paper. However,
the important qualitative effects can be estimated rather
simply.  If the magnetic field produced as a result of
this coupling is strong enough, it will significantly
distort the $r$-mode oscillations so as prevent the
amplification of $r$ modes by gravitational radiation.
If the magnetic field is initially weak, it will be
subsequently amplified and cause the instability to die
out as the star spins down.

\section{Kinematic properties of the linear $r$~modes}
\label{kinematics}

	As mentioned above, a key role in the coupling
between \hbox{$r$-mode} oscillations and the magnetic
field is played by the kinematic properties of the
modes. As we will discuss below, a careful look at the
equations of motion will reveal nonlinear effects which
manifest themselves mostly in secular velocity
fields. Within this Section we will assume that the
\hbox{$r$-mode} oscillations have amplitudes which are
{\it constant} in time, that the nonlinear coupling among
different modes is negligible~\cite{setal01}, and that
there is no magnetic field. The analysis of the
kinematical properties of \hbox{$r$-mode} oscillations in
the case of a time-growing mode amplitude will be
discussed in paper II (cf. Section II.A)

\subsection{The Eulerian perturbation velocity field}
\label{evf}

	We begin our analysis by considering the motion
of fiducial fluid elements on an isobaric surface of a
rotating star experiencing \hbox{$r$-mode}
oscillations. For a Newtonian inviscid star, these are
solutions of the perturbed hydrodynamic equations having
Eulerian velocity perturbations of ``axial
type''~\cite{lom98,fm98}. In an orthonormal basis, and at
first order in the star's unperturbed angular velocity
$\Omega$, such
perturbations may be written as
\begin{equation}
\label{deltav}
\delta {\mathbf v}_1(r,\theta,\phi,t) = 
	\alpha \Omega R \left({r\over R}\right)^\ell 
	{\mathbf Y}^{B}_{\ell\,m} e^{i\sigma t} \ ,
\end{equation}
where $R$ is the radius of the unperturbed star and
$\sigma$ is the frequency of the mode in the inertial
frame. The dimensionless coefficient $\alpha$ describes
the amplitude of the perturbation and we have indicated
with and index ``1'' the velocity perturbations which are
linear in $\alpha$. In Eq.~(\ref{deltav}), ${\mathbf
Y}^{B}_{\ell\,m}$ is the magnetic-type vector spherical
harmonic, and may be defined in terms of the spherical
harmonic functions $Y_{\ell\,m}$ by (see,
e.g.,~\cite{t80})
\begin{equation}
\label{Yblm}
{\mathbf Y}^{B}_{\ell\,m} (\theta,\phi)
	= \frac{1}{\sqrt{\ell(\ell+1)}} 
	\left[r {\nabla} \times(r {\nabla} Y_{\ell\,m})\right]
	\ .
\end{equation} 

	Consider now a frame instantaneously corotating
with the star. In this frame, the differential equations
governing the motion of fiducial fluid elements of an
$\ell=m$ mode\footnote{Hereafter we will always refer to
modes for which $m=\ell$.} in the coordinate basis
$(t,\,r,\,\theta,\,\phi)$ are
\begin{mathletters}
\label{eom}
\begin{eqnarray}
\label{rdot}
& & {\dot r}(t,\theta,\phi) = 0 \ , 
\\ \nonumber \\ 
\label{tdot}
& & {\dot \theta}(t,\theta,\phi) =
	\alpha \Omega \left( \frac{r}{R} \right)^{\ell-1} c_{\ell}
	\sin\theta (\sin\theta)^{\ell-2} 
	\cos \left[\ell\phi +
	 \left(\frac{2\Omega}{\ell+1}\right) t\right]\ ,
\\ \nonumber \\ 
\label{pdot}
& & {\dot \phi}(t,\theta,\phi) =
	- \alpha \Omega \left( \frac{r}{R} \right)^{\ell-1} c_{\ell} 
	\cos\theta (\sin\theta)^{\ell-2} 
	\sin \left[\ell\phi + 
	\left(\frac{2\Omega}{\ell+1}\right) t\right]\ ,
\end{eqnarray}
\end{mathletters}
where
\begin{equation}
\label{c_l}
c_{\ell} \equiv
	\left(-\frac{1}{2}\right)^{\ell}\frac{\ell}{(\ell+1)!}
	\sqrt{\frac{(2\ell+1)!}{\pi}} \ ,
\end{equation}
and the dot refers to the total derivative with respect
to the time coordinate. The angular frequency $\sigma$ of
an \hbox{$r$-mode} in the inertial frame can be related
to its angular frequency $\omega$ in the corotating frame
and the stellar angular velocity. At the lowest order in
$\Omega$, this relation is given by~\cite{pp78,a98}
\begin{equation}
\label{omegas}
\sigma = \omega - \ell \Omega = \frac{2 \ell \Omega}{\ell(\ell+1)}
	- \ell \Omega =
        - {(\ell-1)(\ell+2)\over \ell+1}\Omega \ .
\end{equation}

	Throughout this paper and in the companion paper
II we will restrict ourselves to the study of the
\hbox{$r$-mode} instability in the slow rotation
approximation, retaining only the the lowest order term
in $\Omega$. Nevertheless, we will often consider neutron
stars spinning very near the break-up limit. While this
approximation might be a reasonable one~\cite{lmo99}, it
is important to bear in mind that significant
modifications could appear in our picture of the
\hbox{$r$-mode} instability if the general relativistic
rotational effects are fully taken into account.

\subsection{Nonlinear Motions of Fluid Elements with
constant amplitude}
\label{motf}

	As will become apparent in Section
\ref{evol_eqs}, the linear-order equations (\ref{eom}) do
not generate a significant secular magnetic field since
they lead to unitary strain tensors [cf. Eq.
(\ref{intgrl_indctn})]. However, equations (\ref{eom})
can provide important information about the nonlinear
motions of fluid elements and, in particular, about
whether they lead to a secular drift velocity. When the
nonlinear expressions are not available, in fact, a
rather standard technique~\cite{ll87,l80} allows to
calculate second-order quantities from linear
results. This is an approximation but in some relevant
examples, such as sound waves and shallow water waves,
one finds that the lowest-order {\it nonlinear\/}
corrections to the linear velocity field make no
contribution to the estimated velocity field: i.e. the
drift velocity is given exactly to ${\cal O}(\alpha^2)$
by the linear velocity field.

	We have made use of this technique and obtained
analytical expressions for the values of the velocity
perturbations at second-order in $\alpha$. In particular,
we have expanded the equations of motion in powers of
$\alpha$, averaged over a gyration, and retained only the
lowest-order non-vanishing term (see Appendix~\ref{app_a}
for details). Interestingly, we find that a second-order
in $\alpha$ drift velocity exists in the $\phi-$direction
and the total displacement in $\phi$ from the onset of
the oscillation at $t_0$ up to time $t$ is then found to
be (see Appendix~\ref{app_a})\footnote{At second-order in
$\alpha$ there is no secular motion in the
$\theta-$direction.}
\begin{equation}	
\label{anal_0}
\Delta {\tilde x}^{\phi}(r,t) \equiv
	\int^t_{t_0} \delta v^{\phi}_1(t') dt' =  
	\frac{2}{\ell+1}
	\left(\frac{r}{R}\right)^{\ell-1}\kappa_{\ell}(\theta)
	\int^t_{t_0} \alpha^2(t') \Omega(t') dt' 
	+ {\cal O}(\alpha^3)\ .
\end{equation}
with $t \gg \Delta t$. The matching coefficient
$\kappa_{\ell}(\theta)$ is introduced to relate the
instantaneous and secular velocities and is dependent on
the mode number $\ell$ and on the $\theta$ position on an
isobaric surface of the star. For the $\ell=2$,
$\kappa_2(\theta) \equiv (1/2)^7 (5!/\pi)
(\sin^2\theta-2\cos^2\theta)$ and hence the net
displacement in the azimuthal direction after an
oscillation is approximately $2\pi\alpha$ times the
radius of gyration. The presence of a secular motion in
Eq. (\ref{anal_0}) is indicated by a non-periodic
argument for the integral in the second expression on the
left-hand-side. Note also that, for a star with constant
angular velocity, a net secular azimuthal motion is
obtained {\it both} when the mode's amplitude is constant
[in which case this is ${\cal O}(\alpha^2 t)$] and when
the mode's amplitude is exponentially growing in time [in
which case this is ${\cal O}(\alpha^2_0
\exp(2t/\tau_{_{\rm GW}}))$, where $\alpha_0$ is the
initial mode amplitude and $\tau_{_{\rm GW}}$ is the
timescale for the instability to develop]. As we will
discuss in paper II, this latter case will produce an
even stronger azimuthal drift.

	The azimuthal drift velocity of a given fluid
element is readily calculated from (\ref{anal_0}). In an
orthonormal basis and for the $\ell=2$ mode, this is
\begin{equation}
\label{vdrift}
{\mathbf v}_{\rm d}(r,\theta,t) = \frac{2}{3}\kappa_2(\theta) \alpha^2(t)
\Omega(t) R \left(\frac{r}{R}\right)^2 {\bf e}_{\hat \phi} \ ,
\end{equation}
\noindent where ${\bf e}_{\hat \phi}$ is the unit vector
in the $\phi$ direction. It is worth underlining that the
secular velocity field (\ref{vdrift}) is responsible for
a differential rotation in both the radial and polar
directions.

	It is important to emphasize that using equations
(\ref{eom}) to compute the displacement of an element of
fluid expanding ${\dot \theta}$ and ${\dot \phi}$ in
powers of $\alpha$ is not equivalent to considering
nonlinear effects in the fluid equations. The order in
$\alpha$ to which we compute the fluid displacement given
by the linear velocity field (\ref{eom}) is therefore a
distinct (although related) question. We can compute the
fluid motions predicted by the linear velocity field to
any order we like, but if we compute them to orders
higher than $\alpha$ it does {\it not} mean we have
properly considered nonlinear fluid effects. However,
using as guides analogous fluid-dynamical processes whose
nonlinear behaviour is known, we expect the existence of
a secular drift velocity of ${\cal
O}(\alpha^2)$. Moreover, we expect the drift of fluid
elements given by the velocity field ${\mathbf v}_{\rm
d}$ to be qualitatively correct and perhaps exact to
${\cal O}(\alpha^2)$. After this prediction was first
made~\cite{rls00}, the existence of an ${\cal
O}(\alpha^2)$ drift velocity and differential rotation
has been verified both in analytical simplified
models~\cite{lu01} and, at least qualitatively, through
nonlinear numerical simulations~\cite{sf00,ltv00}. It
should also be noted that the differential rotation
described by Eq.~(\ref{vdrift}) is of kinematic nature
and is set up on the timescale given by the rotation
period of the star. This is much shorter than the
timescale of gravitational wave emission over which the
differential rotation, first proposed by
Spruit~\cite{s98}, is produced.
\begin{figure}[htb]
\begin{center}
\leavevmode
\vbox{
\hspace{0.125truecm}
\psfig{file=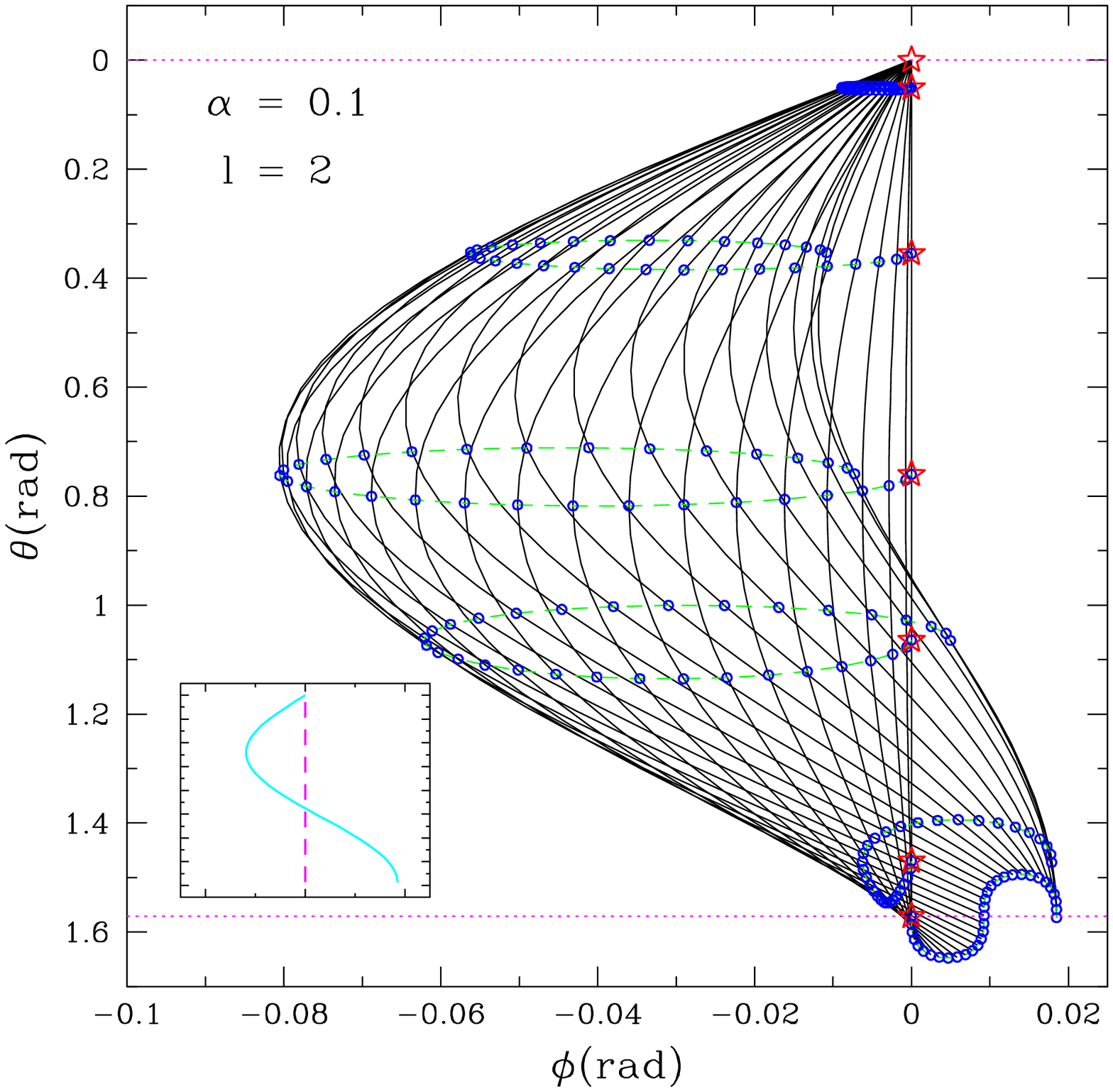,angle=0,width=8.5truecm,height=9.5truecm}
\hspace{0.25truecm}
\psfig{file=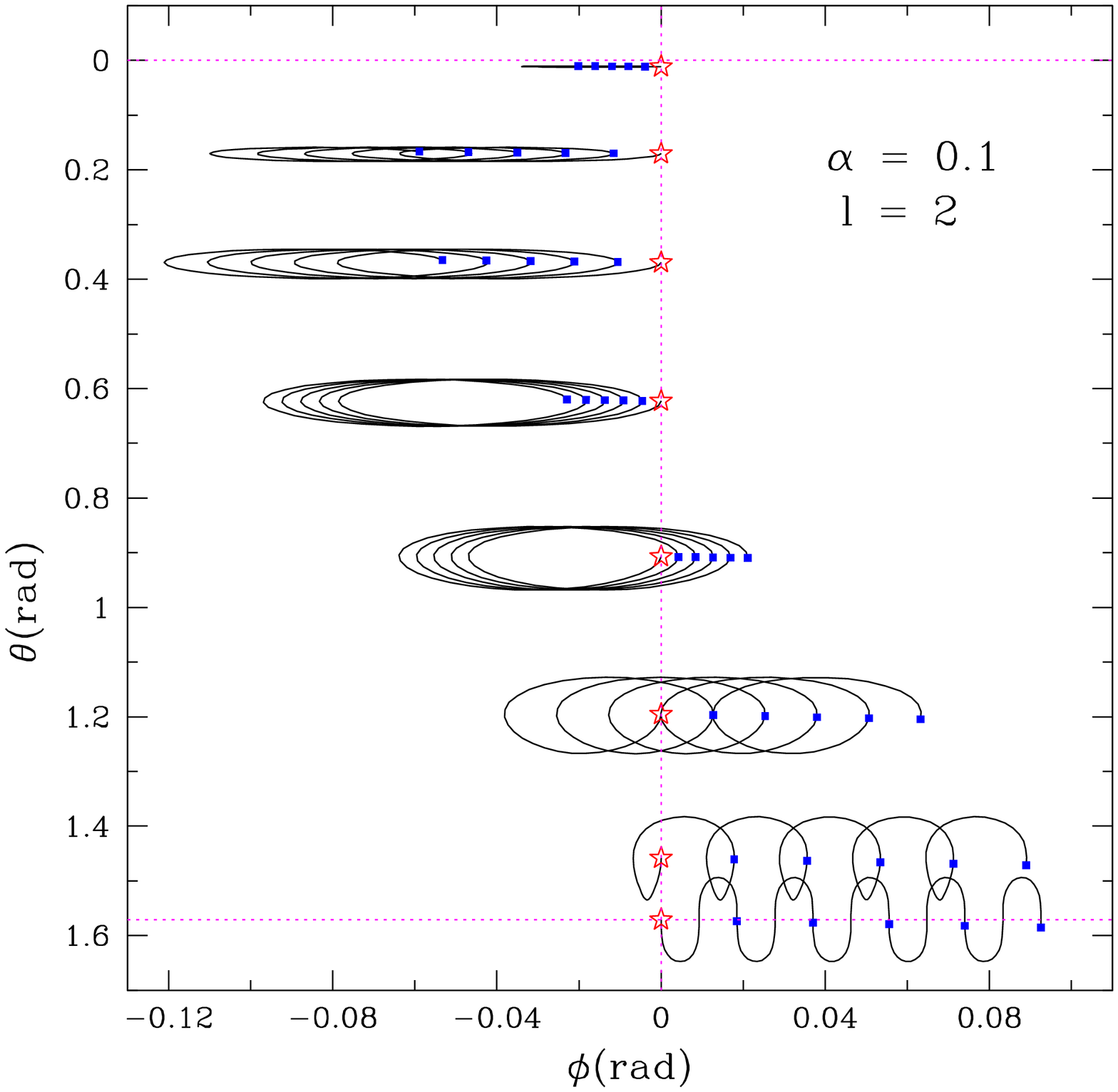,angle=0,width=8.5truecm,height=9.5truecm}
\caption[]{\ Projected trajectories
$\theta(t)\sin\theta(t)\cos\phi(t)$, and
$\phi(t)\sin\theta(t)\cos\phi(t)$ of seven fiducial fluid
elements as seen in the corotating frame. The left
diagram shows the projected trajectories over {\sl one
period} of fiducial fluid elements subject to an $r$-mode
oscillation with $\ell = 2$, $\alpha = 0.1$, and
$\Omega=\Omega_{_B}$. All of the fluid elements are
initially positioned on the $\phi_0=0$ meridian and have
different latitudes (indicated with stars). Continuous
lines show the simultaneous positions of the fluid
elements and different lines refer to different
times. The trajectories of some representative fluid
elements are shown with dashed circles. The small inset
in the left diagram of Fig.~\ref{fig1} shows the initial
positions (dashed line) and final positions (continuous
line) of the different fluid elements. The right panel
uses the same quantities as in the left one, but shows
the trajectories of eight fluid elements for five
oscillations.
\label{fig1}
}}
\end{center}
\end{figure}

	In Fig.~\ref{fig1} we show numerical integrations
of the equations of motion (\ref{tdot}) and (\ref{pdot})
on the northern hemisphere of the rotating star. In
particular, the left panel of Fig.~\ref{fig1} shows the
projected $(\theta, \phi)$ trajectories over one period
for fiducial fluid elements subject to an $r$-mode
oscillation with $\ell = 2$ (the coordinates are those of
a reference frame corotating with the star). All of the
fluid elements are initially positioned on the $\phi_0=0$
meridian, but at different latitudes and these positions
are indicated with stars (because of the polar dependence
of equations [\ref{tdot}] and [\ref{pdot}], it is
sufficient to study them in the northern hemisphere
only). The continuous lines show the simultaneous
positions of fluid elements at different times (some of
these positions are indicated with circles), and the
dashed lines are used to trace the trajectories of
representative fluid elements. In solving equations
(\ref{tdot}) and (\ref{pdot}) we have assumed that
$\alpha$ has the constant value $0.1$ and that the star
is rotating at the break-up limit $\Omega_{_B} \equiv
(2/3)\sqrt{\pi G {\bar \rho}}$, with ${\bar \rho} = 3
M/(4 \pi R^3)$ being the average mass density.

	Several interesting properties can be deduced
from Fig.~\ref{fig1}: {\it a)} the trajectories over one
period $\omega/2\pi$ are {\it not} closed; {\it b)} while
the excursions in the $\theta$ and $\phi$ directions are
comparable, the net displacement $\Delta \phi$ in
longitude after one cycle is much larger than the net
displacement $\Delta \theta$ in latitude (indeed one can
show that this is zero at ${\cal O}(\alpha^2)$); {\it c)}
both $\Delta \phi$ and $\Delta \theta$ are linearly
proportional to $\alpha$; {\it d)} the motions have a
simple dependence on latitude. The small inset in the
left diagram of Fig.~\ref{fig1} shows schematically the
position of the fluid elements at the beginning of the
oscillation (dashed line) and after one cycle (continuous
line). The right panel of Fig.~\ref{fig1} shows the
trajectories of fiducial fluid elements during five full
oscillations and the net displacement produced by their
motions.

	Despite the complexity of the different
trajectories, all of the information about their secular
behaviour is contained in (\ref{anal_0}) and
(\ref{vdrift}). These expressions are at the basis of our
study of the interaction between \hbox{$r$-mode}
oscillations and any magnetic field that is initially
present in the rotating neutron star.

\section{Evolution Equations for the Magnetic Field}
\label{evol_eqs}

	In this Section we first discuss the model
assumed for a magnetic neutron star (Section
\ref{smomns}) and then illustrate the techniques employed
to compute the evolution of an arbitrarily small seed
magnetic field as a result of the secular drift produced
by the \hbox{$r$-mode} oscillations. In particular, we
have considered both Lagrangian and Eulerian formulations
of the induction equation. The first exploits the
possibility of calculating the strain tensor produced by
the oscillation. The second approach, on the other hand,
is more commonly used and treats the induction equation
as a set of partial differential equations. For
compactness, the numerical results obtained using both of
these methods are presented in Paper II.

\subsection{Simplified model of magnetic neutron stars}
\label{smomns}

	The properties of the magnetic fields of neutron
stars still contain a number of unresolved aspects,
mostly connected with the properties of matter at very
high densities~\cite{l91}. For simplicity, we will assume
that the stellar magnetic field ${\mathbf B}$ is
initially dipolar and aligned with the star's spin
axis. Then
\begin{equation}
\label{dipole}
{\mathbf B}_0 = {\mathbf B}^{\hat p}(t=0) =
     B_{_{\rm d}}\frac{R^3}{r^3} \left( 2 \cos\theta
	{\mathbf e}_{\hat {r}} + \sin\theta {\mathbf
	e}_{\hat {\theta}} \right) \ ,
\end{equation}
where $B_{_{\rm d}}$ is the strength of the equatorial
magnetic field at the stellar surface and which is
observed to be in the range $10^{11}-10^{13}$ G. The use
of a dipolar magnetic field allows us to consider the
star as being initially in magnetohydrostatic equilibrium
and therefore avoids the calculation of a stable initial
configuration. Moreover, we will assume that the
electrical currents are concentrated at the origin and to
avoid singularities we restrict our considerations to a
region of the star with radius $pR \le r \le R$, where $0
< p < 1$, and $R$ is the stellar radius. Because any
misalignment between the magnetic field and the rotation
axis will only introduce geometrical corrections
${\mathcal O}(1)$, we expect that all of the features of
the generation and evolution discussed below will not
change significantly when a more generic initial
configuration is considered.

	Except for the very first moments after the
star's birth, the electrons in the outer layers of a
neutron star are strongly degenerate and form an almost
ideal Fermi-gas. Atoms are partially or fully ionized and
form either a strongly coupled Coulomb liquid or a
Coulomb crystal~\cite{yu80}. The electrical and thermal
transport properties of this dense matter are mainly
determined by the transport properties of the electrons,
which are the most important carriers of electrical
charge and heat. At temperatures above the
crystallization temperature of the ions, the electrical
and thermal conductivities are governed by electron
scattering off ions. Increasingly accurate calculations
of the electrical and thermal conductivity of the matter
in hot neutron star envelopes can be found in the
literature~\cite{yu80,ihk93,pbhy99} and an approximate
expression for the electrical conductivity is given
by~\cite{l91}\footnote{Note that expression
(\ref{sigma_E}) is roughly correct for densities in the
range $10^{10}-10^{14}$ g cm$^{-3}$ and temperatures in
the range $10^{6}-10^{8}$ K, but provides a reasonable
estimate also at temperatures of $\sim 10^{9}-10^{10}$ K
which are the relevant ones for the $r$-mode instability
(cf.~\cite{pbhy99}).}
\begin{equation}
\label{sigma_E}
\sigma_{_{E}} \approx 10^{26} \left(\frac{10^8\ \rm K}{T}\right)^2
	\left(\frac{\rho}{10^{14}\ {\rm g\ cm}^{-3}}\right)^{3/4} \
	{\rm s}^{-1}\ ,
\end{equation}
where $T$ and $\rho$ are the stellar temperature and mass
density. Even for a magnetic field that varies on a
length-scale as small as $\Delta R \simeq 0.1 R$, the
magnetic diffusion timescale is
\begin{equation}
\label{tau_diff}
\tau_{\rm diff} = \frac{4\pi (\Delta R)^2 \sigma_{_{E}}}{c^2} \approx 
	3 \times 10^{6} \left(\frac{\Delta R}{10^5\ {\rm cm}}\right)^2
	\left(\frac{10^9\ \rm K}{T}\right)^2
	\left(\frac{\rho}{10^{14}\ {\rm g\ cm}^{-3}}\right)^{3/4} \ 
	{\rm yr}\ .
\end{equation}
This is well over six orders of magnitude larger than the
one year timescale usually discussed for the existence of
of $r$~modes for a typical newly-born neutron star. As a
result, we can neglect the effects of Ohmic dissipation,
treating the fluid as perfectly conducting on the
timescales of interest here.

\subsection{Lagrangian approach}
\label{la}

	In order to simplify our notation, hereafter we
will drop the index ``1'' for the linear velocity so that
$\delta {\mathbf v}\equiv \delta {\mathbf
v_1}$. Combining then the equation of mass conservation
\begin{equation}
\label{mass_cons}
\frac{d \rho }{d t} = - \rho \left( \nabla \cdot {\mathbf v} \right)\ ,
\end{equation}
where $d/dt \equiv (\partial/\partial t + {\mathbf v} \cdot \nabla )$,
with the induction equation
\begin{equation}
\label{induction_0}
\frac{\partial {\mathbf B}}{\partial t}  = 
	\nabla \times \left( 
	{\mathbf v} \times {\mathbf B} \right) \ ,
\end{equation}
one obtains
\begin{equation}
\label{induction_1}
\frac{D }{D t} \left( \frac{{\mathbf B}}{\rho} \right) = 
	\left( \frac{{\mathbf B}}{\rho} \cdot \nabla \right) 
	\delta {\mathbf v}\ ,
\end{equation}
where we have decomposed the velocity ${\mathbf v}$ into
${\mathbf v} = {\mathbf v}_0 + \delta {\mathbf v}$, with
${\mathbf v}_0\equiv \Omega \times {\mathbf r}$ being the
uniform stellar rotation velocity and $\delta {\mathbf
v}$ the \hbox{$r$-mode} velocity perturbation. Here,
${D}/{D t} \equiv ( {\partial}/{\partial t} + {\mathbf v}
\cdot \nabla - {\mathbf \Omega} \times )$ is the
Lagrangian derivative for a fluid element moving at
velocity ${\mathbf v}$ as seen in the corotating
frame. Equation (\ref{induction_1}) can be integrated
analytically to give~\cite{p79,bh98} (see
Appendix~\ref{app_b} for a derivation)
\begin{equation}
\label{intgrl_indctn}
\frac{B^j}{\rho} ({\tilde {\mathbf x}}, t) = 
	\frac {B^k}{\rho} ({\mathbf x}, t_0) 
	\frac{\partial {\tilde x}^j(t)}{\partial x^k(t_0)}\ .
\end{equation}

	Note that $\nabla \cdot {\mathbf v}_0 = 0$ at
first order in the stellar angular velocity and $\nabla
\cdot \delta {\mathbf v} = 0$ by definition of axial
perturbations [cf. equations (\ref{eom})]. To lowest
order in $\Omega$ and $\alpha$ the flow is therefore
incompressible and we can set $\rho ({\tilde {\mathbf
x}}, t) = \rho ({\mathbf x}, t_0)$. The integral form
(\ref{intgrl_indctn}) of the induction equation is
particularly advantageous as it shows that the magnetic
field at time $t$ and position ${\tilde {\mathbf x}}$ can
be computed from the magnetic field at time $t_0$ and
position ${\mathbf x}$, using the tensorial coordinate
strain $\partial {\tilde x}^{j}(t)/\partial {x}^{k}(t_0)$
that develops between $t_0$ and $t$. The advection of
magnetic field lines is an obvious consequence of
Eq.~(\ref{intgrl_indctn}) and the problem of the magnetic
field evolution is therefore transformed into the problem
of determining the time evolution of the strain tensor
$S_{jk}\equiv \partial {\tilde x}^{j}/\partial
{x}^{k}$. While very compact and relatively simpler to
solve numerically, Eq.~(\ref{intgrl_indctn}) has the
disadvantage of being sensitive to the accurate
calculation of the strain tensor, which might become
difficult when the instability is fully developed. For
this reason, and to verify the validity of the Lagrangian
approach for very large saturation amplitudes, we have
also implemented a more traditional Eulerian method for
the solution of the induction equation which is discussed
in the following Section.

\subsection{Orbit Average Eulerian approach}
\label{ea}

	We start by rewriting the induction equation
(\ref{induction_0}) as
\begin{equation}
\label{induction_2}
\frac{{\mathcal D} {\mathbf B}}{{\mathcal D} t} = 
	\nabla \times 
	\left( \delta {\mathbf v} \times {\mathbf B} \right)
	= - \left( \delta {\mathbf v} \cdot \nabla \right){\mathbf B} 
 	+ \left( {\mathbf B} \cdot \nabla \right)\delta {\mathbf v}
	\ , 
\end{equation}
where now ${\mathcal D}{\mathbf B}/{{\mathcal D}t} \equiv
({\partial}/{\partial t} + {\mathbf v}_0 \cdot \nabla -
{\mathbf \Omega} \times) {\mathbf B} = ({D}/{D t} -
\delta{\mathbf v} \cdot \nabla ) {\mathbf B}$ is the time
derivative of the magnetic field in a coordinate system
instantaneously corotating with the star. Writing the
directional derivatives in (\ref{induction_2}) explicitly
within an orthonormal coordinate system $({\mathbf
e}_{\hat {r}}, {\mathbf e}_{\hat {\theta}}, {\mathbf
e}_{\hat {\phi}})$ gives
\begin{mathletters}
\label{ind_eq_f}
\begin{eqnarray}
\label{r_induction_expl}
&&\frac{{\mathcal D} B^{\hat {r}}}{ {\mathcal D} t} = 
	- \delta v^{\hat {r}} \frac{\partial B^{\hat {r}}}{\partial r} 
	+ B^{\hat {r}} \frac{\partial \delta v^{\hat {r}}}{\partial r} 
	+ \frac{B^{\hat {\theta}}}{r}
	\frac{\partial \delta v^{\hat {r}}}{\partial \theta}
	- \frac{\delta v^{\hat {\theta}}}{r}
	\frac{\partial B^{\hat {r}}}{\partial \theta}
	+ \frac{1}{r \sin\theta}
	\left(
	B^{\hat {\phi}}
	\frac{\partial \delta v^{\hat {r}}}{\partial \phi}
	- \delta v^{\hat {\phi}}
	\frac{\partial B^{\hat {r}}}{\partial \phi}
	\right)
	\ ,
\\ \nonumber \\
\label{theta_induction_expl}
&&\frac{{\mathcal D} B^{\hat {\theta}}}{ {\mathcal D} t} = 
	- r \delta v^{\hat {r}} \frac{\partial}{\partial r} 
	\left(\frac{B^{\hat {\theta}}}{r} \right)
	+\, r\, B^{\hat {r}} \frac{\partial}{\partial r} 
	\left(\frac{\delta v^{\hat {\theta}}}{r} \right) 
	- \frac{\delta v^{\hat {\theta}}\sin\theta}{r} 
	\frac{\partial}{\partial \theta}
	\left(\frac{B^{\hat {\theta}}}{\sin\theta}\right) 
	+ \frac{B^{\hat {\theta}}\sin\theta}{r} 
	\frac{\partial}{\partial \theta}
	\left(\frac{\delta v^{\hat {\theta}}}{\sin\theta}\right) 
\nonumber \\ 
&& \hskip 1.4truecm 
	+ \frac{1}{r \sin\theta}\left(
	B^{\hat {\phi}}\frac{\partial
	\delta v^{\hat {\theta}}}{\partial \phi}
	- \delta v^{\hat {\phi}}
	\frac{\partial B^{\hat {\theta}}}{\partial \phi}
	\right)
	\ ,
\\ \nonumber \\
\label{phi_induction_expl}
&&\frac{{\mathcal D} B^{\hat {\phi}}}{ {\mathcal D} t} = 
	- r \delta v^{\hat {r}} \frac{\partial}{\partial r} 
	\left(\frac{B^{\hat {\phi}}}{r} \right)
	+\, r\, B^{\hat {r}} \frac{\partial}{\partial r} 
	\left(\frac{\delta v^{\hat {\phi}}}{r} \right) 
	- \frac{\delta v^{\hat {\theta}}\sin\theta}{r} 
	\frac{\partial}{\partial \theta}
	\left(\frac{B^{\hat {\phi}}}{\sin\theta}\right) 
	+ \frac{B^{\hat {\theta}}\sin\theta}{r} 
	\frac{\partial}{\partial \theta}
	\left(\frac{\delta v^{\hat {\phi}}}{\sin\theta}\right) 
\nonumber \\ 
	&& \hskip 1.4truecm 
	+ \frac{1}{r \sin\theta}\left(
	B^{\hat {\phi}}
	\frac{\partial \delta v^{\hat {\phi}}}{\partial \phi}
	- \delta v^{\hat {\phi}}
	\frac{\partial B^{\hat {\phi}}}{\partial \phi}
	\right)
	\ .
\end{eqnarray}
\end{mathletters}

	The induction equations (\ref{ind_eq_f}) are not
yet in a useful form since they refer to instantaneous
values of the velocity perturbations. The magnetic field
produced over a single oscillation is simply estimated to
be $\delta B \approx \pi \alpha B$ and is uninterestingly
small unless the seed magnetic field is already very
large\footnote{Note that although the newly generated
magnetic field is small, the magnetic tension forces due
to the non-axisymmetric deformation of ${\mathbf B}_0$
might well be comparable with the driving
radiation-reaction force, significantly distorting the
character of the $r$-mode oscillations. See the
discussion in Sect. IV of paper II.}. We need therefore
to perform an orbit average of the induction equations
(\ref{ind_eq_f}) over a timescale $\omega^{-1} \ll \tau
\ll T$, where $T$ is the timescale for a global change of
the velocity and magnetic fields. In this case we can
introduce a time-average (secular) drift velocity
${\tilde {\mathbf v}} \equiv \langle \delta {\mathbf v}
\rangle$, whose components at a time $t$ are defined as
\begin{equation}
\label{sec_vels}
{\tilde v}^{\hat {i}}(r,\theta,\phi,t) \equiv 
	\frac{1}{2 \tau} \int^{\tau}_{-\tau} 
	\delta v^{\hat {i}} (r,\theta,\phi,t+t')\,dt' \ ,
	\hskip 2.0truecm
	i = 2,\ 3. 
\end{equation} 
and where, of course, ${\tilde v}^{\hat {r}}=0$. For
simplicity and because of the smallness of $\delta
{\mathbf B}$, we can neglect the nonlinear term
\hbox{$\nabla \times (\langle \delta {\mathbf v} \times
\delta {\mathbf B} \rangle )$} introduced by the time
average and correlating the velocity perturbation with
the magnetic field perturbation. The orbit average of
equations (\ref{ind_eq_f}) then has the only effect of
replacing $\delta v^{\hat i}$ with ${\tilde v}^{\hat
{i}}$.

	At this point a number of considerations can be
made to simplify the solution of the orbit average of the
Eulerian equations (\ref{ind_eq_f}). Firstly, we note that the
secular velocities must have the same variable
separability as the originating instantaneous velocity
perturbations. As a result, they can be similarly
decomposed as [cf. Eq. (\ref{deltav})]
\begin{equation}
\label{proprty}
{\tilde v^i} = f(r) g_i(\theta,\phi) = r^{\ell} g_i(\theta,\phi) \ , 
\hskip 2.0truecm i = 2,~3 \ ,
\end{equation}
so that $\partial {\tilde v}^{\hat {i}}/\partial r = \ell
{\tilde v}^{\hat {i}}/r$, and $\partial({\tilde v}^{\hat
{i}}/r)/\partial r= (\ell-1) {\tilde v}^{\hat
{i}}/r^2$. By exploiting the property~(\ref{proprty}) 
we are therefore able to remove all of the radial
derivatives in equations (\ref{ind_eq_f}).

\begin{figure}[htb]
\begin{center}
\leavevmode
\psfig{file=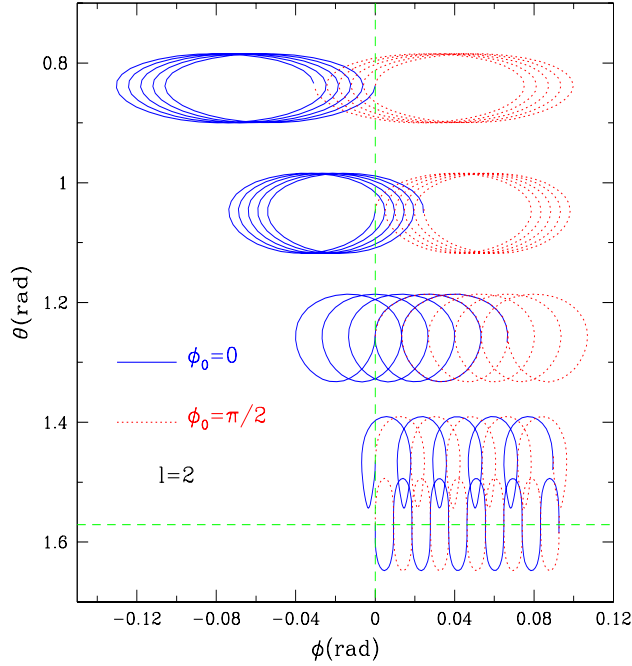,angle=0,width=8.5truecm,height=9.5truecm}
\caption[]{\ This panel shows the trajectories
$\theta(t)$, and $\phi(t)$ of fiducial fluid elements
during five oscillations for a mode with $\ell = 2$,
$\alpha = 0.1$, and $\Omega=\Omega_{_B}$. Continuous
lines refer to fluid elements initially positioned on the
$\phi_0=0$ meridian, while the dotted ones refer to fluid
elements initially positioned on the $\phi_0=\pi/2$
meridian and have been rescaled as $\phi(t) \rightarrow
\phi(t) - \pi/2$ in order to be superimposed on the same
plot.
\label{fig2}}
\end{center}
\end{figure}

	Secondly, while ${\tilde v}^{\hat {\phi}}$ has a
polar dependence, it must be axisymmetric if it is
continuous and periodic. This can most easily be seen by
considering the equations of motion (\ref{eom}). In the
case $\ell=2$, for instance, $\dot \phi$ will be periodic
with period $\pi$ and will change sign at $\phi_0 =
\pi/2$. A similar $\pi$-periodicity must be expected also
for ${\tilde v}^{\hat {\phi}}$ which, however, cannot
change sign at $\phi_0 = \pi/2$, nor vary as a function
$\phi$, since either of these two features would make
fluid elements converge secularly and thus produce
discontinuities.  This property of ${\tilde v}^{\hat
{\phi}}$ is synthesized in Fig.~\ref{fig2}, which shows
the trajectories of fiducial fluid elements during five
oscillations and the overall displacement that results
from these motions (this is the same as the left panel of
Fig.~\ref{fig1}, but here we have considered only fluid
motions near to the star's equator). In particular, we
show fluid trajectories for two different initial
longitudes, i.e. $\phi_0=0$ and $\phi_0=\pi/2$, and
rescale the latter trajectories so that they can be
superimposed on the same plot. As expected, $\dot \phi$
has different signs for the two initial data, but in both
cases ${\tilde v}^{\hat {\phi}}$ has the same sign,
magnitude and polar dependence, so that we can set
\begin{equation}
\label{axs1}
\frac{\partial {\tilde v}^{\hat {\phi}}}{\partial \phi}=0
\ .
\end{equation}

	Thirdly, while ${\tilde v}^{\hat {\theta}}$ does
not have symmetries, it has some important properties:
{\it i)} it has an overall periodic (in time) behaviour,
with period \hbox{$\gg 2\pi/\omega$}; {\it ii)} the
azimuthal dependence is of the form $e^{i m \phi}$; {\it
iii)} it is always much smaller than ${\tilde v}^{\hat
{\phi}}$; {\it iv)} the polar dependence involves only
the lowest order Legendre polynomials.

\begin{figure}[htb]
\begin{center}
\leavevmode
\psfig{file=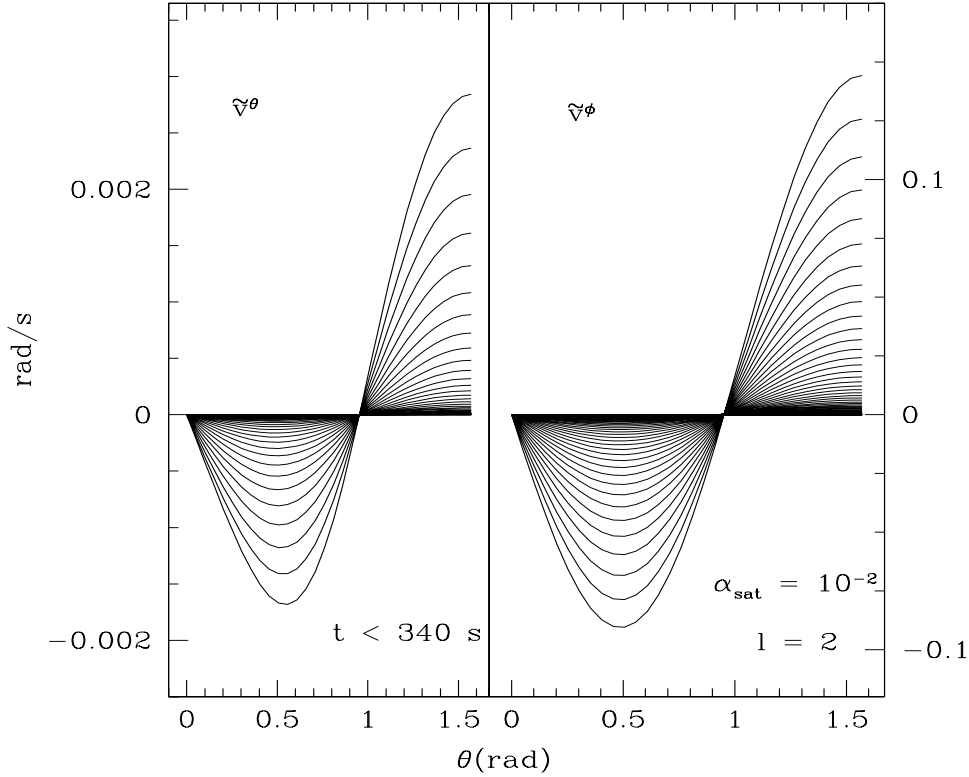,angle=0,width=13.0truecm,height=11.0truecm}
\caption[]{\ Time evolution of the polar profile of
${\tilde v}^{\hat {\theta}}$ and ${\tilde v}^{\hat
{\phi}}$ at the surface of the star and for an $\ell = 2$
mode, saturated at an amplitude $\alpha_{\rm sat} =
10^{-2}$. Different curves refer to different times and
show the velocity increase produced by the growth of the
mode amplitude. The velocity profiles shown here refer to
a time \hbox{$t < 340$ s}. The velocity and subsequently
decreases as a result of the star's spin down (not shown
here).
\label{fig3}}
\end{center}
\end{figure}
This periodic behaviour of ${\tilde v}^{\hat {\theta}}$
guarantees that the polar deformations of the magnetic
field are confined to very small scales and, when
averaged over the relevant timescales, will produce
negligible net effects.

	Properties {\it iii)} and {\it iv)} are shown in
Fig.~\ref{fig3} where we have plotted the evolution of
the polar profiles for ${\tilde v}^{\hat {\theta}}$ and
${\tilde v}^{\hat {\phi}}$ at the surface of the star
during the growth of an $\ell = 2$ mode. Different curves
refer to different times and show the progressive
increase of ${\tilde v}^{\hat {\theta}}$ and ${\tilde
v}^{\hat {\phi}}$ as a result of the mode's growth. The
largest velocity values are reached at saturation (which
was here set to be $\alpha_{\rm sat} = 10^{-2}$ and
occurs at $t\simeq 340$ s) and it should be noted that
${\tilde v}^{\hat {\phi}}$ is almost two orders of
magnitude larger than ${\tilde v}^{\hat {\theta}}$. Since
both ${\tilde v}^{\hat {\theta}}$ and ${\tilde v}^{\hat
{\phi}}$ depend linearly on $\Omega$ [cf. equations
[\ref{eom})], their values progressively decrease after
saturation as a result of the star's spin-down (This is
not shown in Fig.~\ref{fig3}.).

	The azimuthal dependence of ${\tilde v}^{\hat
{\theta}}$ is the same that would be imprinted on the
magnetic field produced by the poloidal velocities. In
fact, even an initially axisymmetric magnetic field would
acquire non-axisymmetric features driven by ${\tilde
v}^{\hat {\theta}}$. However, because these departures
away from axisymmetry are always linearly dependent on
${\tilde v}^{\hat {\theta}}$ and on its
$\phi$-derivative, we will assume them to be negligible
and set
\begin{equation}
\label{axs2}
\frac{\partial {\tilde v}^{\hat {\theta}}}{\partial \phi} 
	= \frac{\partial B^{\hat {r}}}{\partial \phi} 
	= \frac{\partial B^{\hat {\theta}}}{\partial \phi}
	= \frac{\partial B^{\hat {\phi}}}{\partial \phi}=0
	\ .
\end{equation} 
As a result of (\ref{axs1}), (\ref{axs2}) and after some
regrouping, the induction equations (\ref{ind_eq_f}) take
the form
\begin{mathletters}
\label{ind_eq_r}
\begin{eqnarray}
\label{r_induction_fin}
&&\frac{{\mathcal D} B^{\hat {r}}}{{\mathcal D} t} = 
	- \frac{{\tilde v}^{\hat {\theta}}}{r}
	\frac{\partial B^{\hat {r}}}{\partial \theta}
	\ ,
\\ \nonumber \\ 
\label{theta_induction_fin}
&&\frac{{\mathcal D} B^{\hat {\theta}}}{{\mathcal D} t} = 
	\frac{B^{\hat {r}}{\tilde v}^{\hat {\theta}}}{r}
	- \frac{{\tilde v}^{\hat {\theta}}}{r}
	\frac{\partial B^{\hat {\theta}}}{\partial \theta} 
	+ \frac{B^{\hat {\theta}}}{r}
	\frac{\partial {\tilde v}^{\hat {\theta}}}{\partial \theta} 
	\ ,
\\ \nonumber \\ 
\label{phi_induction_fin}
&&\frac{{\mathcal D} B^{\hat {\phi}}}{{\mathcal D} t} = 
	\frac{B^{\hat {r}} {\tilde v}^{\hat {\phi}}}{r}
	- \frac{{\tilde v}^{\hat {\theta}}}{r}
	\frac{\partial B^{\hat {\phi}}}{\partial \theta} 
	+ \frac{{B}^{\hat {\theta}}}{r}
	\frac{\partial {\tilde v}^{\hat {\phi}}}{\partial \theta} 
	+ \left({\tilde v}^{\hat {\theta}} B^{\hat {\phi}}
	- B^{\hat {\theta}}{\tilde v}^{\hat {\phi}}\right)
	\frac{\cot \theta}{r}
	\ ,
\end{eqnarray}
\end{mathletters}
\noindent The set of equations (\ref{ind_eq_r}) has now
only time and polar derivatives, and can be more easily
solved numerically. Results of the numerical integration
of (\ref{ind_eq_r}) and of the comparison with results
obtained from the Lagrangian approach will be presented
in paper II. It is important to underline that the only
approximation made in obtaining equations
(\ref{ind_eq_r}) from the more general equations
(\ref{ind_eq_f}) comes from ignoring the deviations away
from axisymmetry in the velocity field. In this respect,
dynamo theory and in particular Braginsky's dynamo
model~\cite{b64}, suggests that such an assumption can
only lead to an underestimation of the actual growth rate
of the magnetic field~\cite{m78,p79}.
	
	The Eulerian or orbit average approach presented
above is clearly more complicated to implement
numerically than the corresponding Lagrangian method as
it involves the use of a numerical grid on which the set
of coupled partial differential equations needs to be
solved. However, the Eulerian method has also been shown
to be more accurate when the \hbox{$r$-mode} instability
has reached saturation and the velocity perturbations at
the star's surface are ${\cal O}(\Omega R)$. In paper II
we will present a close comparison of the two approaches
discussed above.

\subsection{Simplified analytical model of the r-mode instability}
\label{amrw}

	While a detailed discussion of the full numerical
calculations will be presented in paper II, in what
follows we show how, with simple back of the envelope
calculations, it is possible to predict the generation of
very large toroidal magnetic fields as a result of the
azimuthal secular velocity field produced by the onset
and saturation of the $r$-mode instability.  The
estimates discussed below will then be confirmed by the
full numerical calculations.

	We start by considering the phenomenological
expressions for the time evolution of the stellar angular
velocity and the mode amplitude during the period of
activity of the instability~\cite{oetal98}. These can be
summarized analytically as (see also Fig.~1 of paper~II)
\begin{eqnarray}
\label{anal_2}
\Omega(t) & = & \left\{
	\begin{array}{ll}
	\Omega_0  & 
	\hskip 1.1truecm \textrm{~if \  $t \leq 
	t_{\rm sat}$}\ ,
	\\ 
	\Omega_0 \left[
	1 + C_1 \Omega^{2\ell+2}_0 (t - t_{\rm sat})
	\right]^{-(2\ell+2)} & 
	\hskip 1.1truecm \textrm{~if \  $t > 
	t_{\rm sat}$} \ ,
	\end{array} \right. 
\\ \nonumber \\ 
\label{anal_3}
\alpha(t) &=& \left\{
	\begin{array}{ll}
	\alpha_0 \; \textrm{exp} ( t/|\tau_{_{\rm GW}}|)
	& \hskip 4.0truecm \textrm{if \  $t \leq 
	t_{\rm sat}$}\ ,
	\\
	\alpha_{\rm sat}
	& \hskip 4.0truecm \textrm{if \  $t > 
	t_{\rm sat}$}
	\ .
	\end{array} \right. 
\\ \nonumber
\end{eqnarray}
Here $|\tau_{_{\rm GW}}|$ is the timescale for the onset
of the instability, while $\Omega_0$ is the star's
initial angular velocity. $C_1$ is a short-hand for
\begin{equation}
C_1 \equiv
	\frac{64\pi G}{c^{2\ell+3}}
	\frac{(2\ell+2)(\ell-1)^{2\ell}}{[(2\ell+1)!!]^{2}}
	\left( \frac{\ell+2}{\ell+1} \right)^{2\ell+2}
	\frac{Q \alpha^2_{\rm sat}}{1-Q \alpha^2_{\rm sat} }
	\int_0^R\rho \,r^{2 \ell+2} dr \ , 
\end{equation}
with $Q$ being a nondimensional quantity accounting for
the internal structure of the star. For an $\ell = m = 2$
mode~\cite{oetal98}
\begin{equation}
Q \equiv \frac{9}{16 \pi R^2}
	\left({\int^R_0 \rho \, r^6 dr}\right)
	\left({\int^R_0 \rho \, r^4 dr}\right)^{-1} \ .
\end{equation}	

	Using (\ref{anal_2}) and (\ref{anal_3}),
Eq.~(\ref{anal_0}) can be integrated to give
\begin{eqnarray}
\label{phit_e}
&& {\tilde x}^{\phi}(t) = \left\{
	\begin{array}{ll}
\displaystyle
	\left(\frac{r}{R}\right)
	\frac{\kappa_2\Omega_0 \alpha^2_0}{2}
	\, |\tau_{_{\rm GW}}|
	\biggl[\textrm{exp}( 2t /|\tau_{_{\rm GW}}|) - 1\biggr] 
	\ , &
	\hskip 0.75 truecm \textrm{if \  $t \leq t_{\rm sat}$}\ , 
	\\ \nonumber \\ 
\displaystyle
	\left(\frac{r}{R}\right)
	\frac{\kappa_2\Omega_0 \alpha^2_0}{2}
	\, |\tau_{_{\rm GW}}|
	\biggl[\textrm{exp}( 2t /|\tau_{_{\rm GW}}|) - 1\biggr] \ + \ 
	\\ 
\displaystyle
	\hskip 2.0truecm \left(\frac{6r}{5R}\right)
	\frac{(\kappa_2)^2\alpha^2_{\rm sat}}{C_1 \Omega^5_0} 
	\biggl\{ \left[1 + 
	C_1 \Omega^6_0 (t - t_{\rm sat})\right]^{5/6} - 1 
	\biggr\}\ ,
	&
	\hskip 0.75 truecm \textrm{if \  $t > t_{\rm sat}$}\ . 
	\end{array} \right. \nonumber \\
\end{eqnarray}

	Using Eq.~(\ref{intgrl_indctn}), we can now
estimate the average magnetic field produced at a given
time at the surface of the star as a result of the
secular drift (\ref{phit_e}). Assuming the initial
magnetic field to be predominantly poloidal,
i.e. $B^{\phi}({\mathbf x}, t_0) \ll B^{\theta}({\mathbf
x}, t_0)$, the toroidal magnetic field produced is then
\begin{equation}
\label{anal_1}
B^{\phi} ({\tilde {\mathbf x}}, t) = 
	B^r ({\mathbf x}, t_0) \frac{{\tilde x}^{\phi}(t)}{r}+
	B^{\theta} ({\mathbf x}, t_0) 
	\frac{\partial{\tilde x}^{\phi}(t)}{\partial\theta}
	\ .
\end{equation}
For a ``fiducial'' neutron star, with mass $M=1.4 {\rm
M}_\odot$, radius $R = 12.5$ km, and initial angular
velocity $\Omega_0 $ equal to $\Omega_{_B}$, the
timescale for the onset of an unstable $\ell = m = 2$
mode is then $|\tau_{_{\rm GW}}| \simeq 37$ s. If the
initial perturbation has amplitude $\alpha_0 = 10^{-6}$,
the mode will saturate at $\alpha_{\rm sat}=0.1$ after a
time $t_{\rm sat} \simeq 430$ s [cf. Eq. (\ref{anal_3})]
and the volume averaged toroidal magnetic fields at
saturation and after one year can be estimated to be
respectively
\begin{eqnarray}
\label{intgrl_indctn_phi_e}
&& \langle {\Delta B}^{\hat \phi}(t = t_{\rm sat}) \rangle \simeq 
	\int_{\rm V_*} 
	\left[{B}^{\hat r}_0 + \frac{2}{\pi}
	{B}^{\hat \theta}_0 \right] {\tilde x^{\hat \phi}} 
	{\;d^3{\bf x}}
	\sim 3.9 \times 10^2 \ \langle {B}^{\hat {p}}_0 \rangle 
	\ {\rm G}\ , 
\nonumber \\ \nonumber \\
&& \langle {\Delta B}^{\hat \phi}(t = 1 {\rm yr}) \rangle 
	\sim 1.2 \times 10^8 \ \langle {B}^{\hat {p}}_0 \rangle
	\ {\rm G}\ ,
\end{eqnarray}
where $ \langle ({B}^{\hat {p}}_0)^2 \rangle \equiv
\langle ({B}^{\hat r}_0)^2 + ({B}^{\hat \theta}_0)^2
\rangle$ is the initial poloidal magnetic field averaged
over the stellar volume ${\rm V_*}$. Expressions
(\ref{intgrl_indctn_phi_e}) show that, in the absence of
a back-reaction on the kinematics of the $r$-mode
instability, the toroidal magnetic field is tightly
wrapped around the star so as to have become about two
orders of magnitude larger that the seed poloidal
magnetic field in the short time necessary for the
instability to reach saturation.  Moreover, the total
magnetic field could be amplified by eight orders of
magnitude on a timescale of one year.

	More detailed computations of the evolution of
the magnetic field will be presented in paper
II. However, the simple estimates outlined above already
show that an initial magnetic field ${B}^{\hat {p}}_0$
could produce, after one year, an equipartition toroidal
magnetic field of $\sim 10^{15}$ G, i.e. a toroidal
magnetic field whose energy is comparable with the
rotational energy of the star. Thus, an initial magnetic
field exceeding $10^8$ G, (much below the measured
magnetic field in young pulsars) would cause significant
departures from the standard evolution of the $r$-mode
instability. The impact of these very intense magnetic
fields on the existence or growth of the $r$-mode
oscillations will be discussed in detail in paper
II. There, it will be possible to calculate the strength
of the magnetic field necessary to significantly {\it
distort} the first \hbox{$r$-mode} oscillation, or {\it
suppress} the instability when this is free to develop.

\section{Conclusions}
\label{conclusions}

	We have investigated the onset and growth of the
\hbox{$r$-mode} instability in rotating magnetized
neutron stars. Because of the high conductivity of the
hot neutron star matter and the peculiar nature of the
instability which is powered by large scale mass
currents, it is not possible to ignore the presence of
the strong magnetic fields that are expected to accompany
newly born neutron stars.
	
	Expanding the perturbed velocity field in powers
of the mode amplitude we have derived a second-order
analytic expression for a secular velocity field which we
expect to emerge during the nonlinear growth of the
instability. These secular motions produce a differential
rotation both in the radial and in the polar
directions. On an isobaric surface, the secular effects
manifest themselves as a toroidal drift and couple with
any pre-existing magnetic field to produce toroidal
magnetic fields that rapidly grow in magnitude. In order
to study how these kinematic effects interact with a
magnetic field, we have discussed two different
approaches to the solution of the induction equation and
have derived sets of equations for the two cases. While
we have left the discussion of the numerical results
obtained to the companion paper II, we have here provided
first estimates of the magnitudes of the magnetic fields
that would be produced as a result of the shearing of a
pre-existing poloidal magnetic field into a toroidal
one. In particular, we have shown that it could be
relatively simple to obtain, on the timescale usually
discussed for the existence of the \hbox{$r$-mode}
instability, magnetic fields that would be in
equipartition the rotational kinetic energy of the
star. The magnetic fields that are produced in this way
will influence the \hbox{$r$-mode} instability either by
preventing its onset (when sufficiently strong) or by
suppressing its saturated development. Precise estimates
of the critical magnetic field for prevention and damping
of the instability will be presented in paper II.

\acknowledgments We are grateful to Nils Andersson,
Kostas Kokkotas, John Miller, Nikolaos Stergioulas and
Shin'ichirou Yoshida for the useful discussions and for
carefully reading the manuscript. LR acknowlodges support
from the Italian MURST and by the EU Programme ``Improving
the Human Research Potential and the Socio-Economic
Knowledge Base'' (Research Training Network Contract
HPRN-CT-2000-00137). FKL, DM and SLS acknowledge support
from the NSF grants AST~96-18524 and PHY~99-02833 and
NASA grants NAG~5-8424 and NAG~5-7152 at Illinois. FKL is
also grateful for the hospitality extended to him by John
Miller, SISSA, and ICTP, where this work was completed.

\appendix

\section{Equations of Motion: nonlinear effects from
	linearized equations}
\label{app_a}

	In this Appendix we outline the perturbative
analysis which allows the periodic and secular parts in
the linearized equations (\ref{eom}) to be
distinguished. Here we will assume $\ell=2$, but the
formalism presented can easily be generalized to the case
of an arbitrary $\ell$. We start by rewriting (\ref{eom})
as
\begin{mathletters}
\label{eom_appndx}
\begin{eqnarray}
\label{tdot_appndx}
& & {\dot \theta}(t,\theta,\phi) = \alpha A
	\sin\theta \cos \left(  2\phi +  \omega t \right)\ , 
\\ \nonumber \\
\label{pdot_appndx}
& & {\dot \phi}(t,\theta,\phi) = -\alpha A
	\cos\theta \sin \left(  2\phi +  \omega t \right)\ ,
\end{eqnarray}
\end{mathletters}
where 
\begin{equation}
\label{a_const}
A \equiv \omega \left( \frac{r}{R} \right) c_{2} =
	 \omega \left( \frac{r}{R} \right) 
	\frac{1}{8} \sqrt{\frac{5!}{\pi}} \ .
\end{equation}

	Next, we expand the solution of
(\ref{eom_appndx}) in a series of powers of the mode
amplitude $\alpha$, i.e. we look for solutions of the
form
\begin{mathletters}
\label{eom_pwrs}
\begin{eqnarray}
\label{tdot_pwrs}
& & {\theta}(t) = 
	\phi_0 + \alpha \phi_1(t) + \alpha^2 \phi_2(t) 
	+ {\mathcal O}(\alpha^3) \ ,
\\ \nonumber \\
\label{pdot_pwrs}
& & {\phi}(t) = 
	\theta_0 + \alpha \theta_1(t) + \alpha^2 \theta_2(t) +
	{\mathcal O}(\alpha^3) \ .
\end{eqnarray}
\end{mathletters}
	
	Substituting (\ref{eom_pwrs}) into (\ref{eom_appndx}) yields
\begin{mathletters}
\label{eom_expndd}
\begin{eqnarray}
\label{tdot_expndd}
& & {\dot \theta}(t,\theta,\phi) = \alpha A
	\bigg\{\left[\sin\theta_0 \cos(\theta_1 + \theta_2) + 
	\sin(\theta_1 + \theta_2) \cos\theta_0\right] \times
\nonumber \\
	&& \hskip 5.0truecm
	\left[
	\cos (2\phi_0 +  \omega t )\cos (2\phi_1 + 2\phi_2) - 
	\sin (2\phi_0 +  \omega t )\sin (2\phi_1 + 2\phi_2)
	\right]\bigg\}	\ , 
\\ \nonumber \\
\label{pdot_expndd}
& & {\dot \phi}(t,\theta,\phi) = -\alpha A
	\bigg\{\left[\cos\theta_0 \cos(\theta_1 + \theta_2) -
	\sin(\theta_1 + \theta_2) \sin\theta_0\right] \times 
\nonumber \\
	&& \hskip 5.0truecm
	\left[
	\sin (2\phi_0 +  \omega t )\cos (2\phi_1 + 2\phi_2) +
	\cos (2\phi_0 +  \omega t )\sin (2\phi_1 + 2\phi_2)
	\right]\bigg\}	\ . 
\end{eqnarray}
\end{mathletters}
where we have used the abbreviated notation of $\theta_i$
for $\alpha^i \theta_i$ and $\phi_i$ for $\alpha^i
\phi_i$. Making use of the relations
\begin{eqnarray}
&& \cos(\theta_1 + \theta_2) = 1 - \frac{\theta_1}{2} + 
	{\mathcal O}(\alpha^3) \ , 
\\\nonumber \\
&& \sin (\theta_1 + \theta_2) = \theta_1 + \theta_2 + 
	{\mathcal O}(\alpha^3) \ , 
\\\nonumber \\
&& \cos(2\phi_1 + 2\phi_2) = 1 - 2\phi^2_1 + 
	{\mathcal O}(\alpha^3) \ , 
\\\nonumber \\
&& \sin(2\phi_1 + 2\phi_2) = 2(\phi_1 + \phi_2) + 
	{\mathcal O}(\alpha^3) \ , 
\end{eqnarray}
equations (\ref{eom_expndd}) take the form
\begin{mathletters}
\label{eom_grp}
\begin{eqnarray}
\label{tdot_grp}
& & {\dot \theta}(t,\theta,\phi) = 
	{\dot \theta_0} + {\dot \theta_1} + {\dot \theta_2} + 
	{\dot \theta_3} + {\mathcal O}(\alpha^4) \ ,
\\\nonumber \\
\label{pdot_grp}
& & {\dot \phi}(t,\theta,\phi) =  
	{\dot \phi_0} + {\dot \phi_1} + {\dot \phi_2} + 
	{\dot \phi_3} + {\mathcal O}(\alpha^4) \ ,
\end{eqnarray}
\end{mathletters}
where
\begin{mathletters}
\label{tdots}
\begin{eqnarray}
\label{tdot_0}
& & \hskip 1.0truecm{\dot \theta}_0(t) = 0 \ ,
\\\nonumber \\
\label{tdot_1}
& & \hskip 1.0truecm{\dot \theta}_1(t) = \alpha A \sin\theta_0 
	\cos(2\phi_0 + \omega t) \ ,
\\\nonumber \\
\label{tdot_2}
& & \hskip 1.0truecm{\dot \theta}_2(t) = \alpha A \left[
	\theta_1 \cos\theta_0 \cos(2\phi_0 + \omega t)  -
	2 \phi_1 \sin\theta_0 \sin(2\phi_0 + \omega t) \right]
\ ,
\end{eqnarray}
\end{mathletters}
and
\begin{mathletters}
\label{pdots}
\begin{eqnarray}
\label{pdot_0}
& & \hskip -3.0truecm {\dot \phi}_0(t) = 0 \ ,
\\\nonumber \\
\label{pdot_1}
& & \hskip -3.0truecm {\dot \phi}_1(t) = - \alpha A \cos\theta_0 
	\sin(2\phi_0 + \omega t) \ ,
\\\nonumber \\
\label{pdot_2}
& & \hskip -3.0truecm {\dot \phi}_2(t) =  \alpha A \left[
	\theta_1 \sin\theta_0 \sin(2\phi_0 + \omega t)  -
	2 \phi_1 \cos\theta_0 \cos(2\phi_0 + \omega t) \right] \ \ .
\end{eqnarray}
\end{mathletters}
	
	Note that, in principle, both $\alpha$ and
$\omega$ (and therefore $A$) have a dependence on time
which at first order in $\alpha$ produces orbits with
increasing radius of gyration. Considering both $\alpha$
and $\omega$ as constant, we can easily find the
integrals of equations (\ref{tdots}) through an iterative
procedure.  In particular, we substitute into the
right-hand-side of the equations for ${\dot \theta_i}(t)$
and ${\dot \phi_i}(t)$, the integrals of the equations
for ${\dot \theta_{i-1}}(t)$ and ${\dot
\phi_{i-1}}(t)$. By doing so we then obtain the following
equations of motions
\begin{mathletters}
\label{ts}
\begin{eqnarray}
\label{t_0}
& & \hskip -3.5truecm{\ \theta}_0 = {\rm const.} \ ,
\\\nonumber \\
\label{t_1}
& & \hskip -3.5truecm{\ \theta}_1(t) = \frac{\alpha A}{\omega} \sin\theta_0 
	\sin(2\phi_0 + \omega t) \ ,
\\\nonumber \\
\label{t_2}
& & \hskip -3.5truecm{\ \theta}_2(t) = -\frac{\alpha^2 A^2}{2 \omega^2} 
	\sin\theta_0 \cos\theta_0 \sin^2(2\phi_0 + \omega t)
	\ ,
\end{eqnarray}
\end{mathletters}
and
\begin{mathletters}
\label{ps}
\begin{eqnarray}
\label{p_0}
& & \hskip -1.0truecm {\ \phi}_0 = {\rm const.} \ ,
\\\nonumber \\
\label{p_1}
& & \hskip -1.0truecm {\ \phi}_1(t) = \frac{\alpha A}{\omega} \cos\theta_0 
	\cos(2\phi_0 + \omega t) \ ,
\\\nonumber \\
\label{p_2}
& & \hskip -1.0truecm {\ \phi}_2(t) = -\frac{\alpha^2 A^2}{2 \omega^2} 
	\left[
	(2\phi_0 + \omega t)
	\left(2\cos^2\theta_0 - \sin^2\theta_0\right) + 
	\sin(2\phi_0 + \omega t)\cos(2\phi_0 + \omega t)
	\left(2\cos^2\theta_0 + \sin^2\theta_0\right)
	\right] \ , 
\\\nonumber \\
\label{p_3}
& & \hskip -1.0truecm {\ \phi}_3(t) = \alpha A \Bigg[
	2 \phi^2_1 \cos\theta_0 \sin(2\phi_0 + \omega t) -
	2 \phi_2   \cos\theta_0 \cos(2\phi_0 + \omega t) +
\nonumber \\
& & \hskip 4.0truecm
	2 \theta_1 \phi_1 \sin\theta_0 \cos(2\phi_0 + \omega t) +
	\theta_2 \sin\theta_0 \sin(2\phi_0 + \omega t) \Bigg]
	\ .
\end{eqnarray}
\end{mathletters}
A rapid look at equations (\ref{ts}) and (\ref{ps}) shows
that while ${\theta}_1(t)$, ${\theta}_2(t)$, and ${\
\phi}_1(t)$ are periodic functions, ${\theta}_3(t)$ and
${\ \phi}_2(t)$ have terms that grow linearly in time and
are responsible for an ${\mathcal O}(\alpha^3)$ drift in
latitude and an ${\mathcal O}(\alpha^2)$ drift in
longitude.

	As discussed in the main text, using equations
(\ref{eom}) to compute the displacement of an element of
fluid by expanding ${\dot \theta}$ and ${\dot \phi}$ in
powers of $\alpha$ is not equivalent to considering
nonlinear effects in the fluid equations. The key issue
is whether the fluid drift given by (\ref{eom}) is at
least qualitatively correct. In principle it might not
be, because the velocity field obtained by solving the
nonlinear fluid equations that describe $r$~waves might
have additional parts that contribute to the drift and
largely or completely cancel the drift given by the
linear velocity field. Nonlinear solutions for the
$r$~waves are not yet available, so it is natural to ask
how the motions of fluid elements that we have found for
the $r$~waves compare with the motions of fluid elements
in other, more familiar waves.

	Sound waves cause fluid elements to drift as well
as to oscillate, as discussed by Landau and
Lifshitz~\cite{ll87}, who show how to compute the mass
current density produced by a sound wave in the Eulerian
frame, using the solution of the linearized fluid
equations. The drift is of second-order in the wave
amplitude. Calculating the fluid drift produced by sound
waves as we have computed the drift produced by
$r$~waves, i.e., by using the velocity field (\ref{eom})
in the Lagrangian frame to compute the motions of
individual fluid elements, gives the drift quoted by
Landau and Lifshitz. Shallow water waves are another
interesting example.  Here we outline how the fluid
equations can be solved exactly for such waves, using the
method of characteristics, and describe the exact
solution. Consider an infinite train of shallow water
waves with wavelength $2\pi k^{-1}$ much larger than the
depth $h_o$ of the unperturbed water. The evolution of
surface waves is then determined by the momentum and mass
conservation equations
\begin{equation}
\frac{\partial u}{\partial t} + 
	u \frac{\partial u}{\partial x}=
	-g \frac{\partial h}{\partial x} 
	\qquad{\rm and}\qquad 
\frac{\partial h}{\partial t} + 
	\frac{\partial (hu)}{\partial x} = 0\ ,
\end{equation}
where $u$ is the horizontal velocity and $g$ is the
gravitational acceleration. Using $c\equiv \sqrt{gh}$
instead of $h$, implicit solutions of arbitrary amplitude
can be expressed in terms of the Riemann invariants,
$I_{\pm}$, of the rightward-moving ($+$) characteristics
and the leftward-moving ($-$) characteristics:
$(dx/dt)_{\pm} = u \pm c$. Setting up at $t=0$ a
right-propagating sinusoidal wave of small amplitude $y =
\alpha c_o \cos(kx)$, $u = 2\alpha c_o \cos(kx)$, where
$y \equiv c - c_o$ and $c_o \equiv \sqrt{gh_o}$ is the
speed of small-amplitude waves, we obtain for the
right-moving characteristics $x(x_o,t) = x_o + c_o [1 +
3\alpha\cos(kx_o)]t$. As the wave propagates, the leading
slopes of the crests steepen and form singularities at
the earliest crossing of the characteristics, which
occurs at $t_{\rm s} = 1/3\alpha c_o k$. At early times
($t\ll t_{\rm s}$), we can use the above characteristic
trajectory to obtain the velocity field to ${\cal
O}\left(\alpha^2\right)$ from the exact solution. The
result is $u = 2\alpha c_o \cos k(x-c_o t) + 3 \alpha^2
tk c_o^2 \sin 2k(x-c_o t)$.  Averaged over the period $T
= 2\pi/k c_o$, the first term gives the drift velocity
$v_{\rm d} = 2\alpha^2 c_o$; the second term describes
oscillations at twice the wave frequency, with the
linearly growing amplitude $(3/2) v_{\rm d} c_o
kt$. Subtracting constant-amplitude oscillations, the
trajectory of a fluid element that is initially at $x_o$
is then $\bar{x} = x_o + v_{\rm d} t + (3/4) v_{\rm d} t
\cos 2k(x_o - c_o t)$. Calculating the fluid drift
produced by this wave in the same way we have done for
the $r$~waves, i.e., using the velocity field
$\delta{\vec v}_1$ in the Lagrangian frame to compute the
motions of individual fluid elements, gives the same
drift velocity as the exact solution, to second order in
$\alpha$.

	This calculation shows that shallow water waves
produce a fluid drift and that the drift obtained by
using the linear velocity field to compute the motion of
an element of fluid is correct to second order in
$\alpha$. Stated differently, when the full fluid
equations that describe shallow water waves are solved
exactly, one finds that the lowest-order {\it
nonlinear\/} corrections to the linear velocity field,
which includes all effects of ${\cal O}(\alpha^2)$, make no
contribution to the drift; the drift is given exactly to
${\cal O}(\alpha^2)$ by the linear velocity field.

	This result is so important that it may be worth
restating it in a different way. Suppose that we solve
the fluid equations for sound waves and shallow water
waves exactly to ${\cal O}(\alpha^2)$ and write the
solution for the Eulerian velocity field as ${\vec
u}(x,t) = \alpha{\vec u}_1 + \alpha^2 {\vec u}_2$. When
evaluated at a fixed location, neither ${\vec u}_1$ nor
${\vec u}_2$ give a secular drift, i.e., a displacement
that is linearly proportional to $t$. ${\vec u}_1$ is a
pure sinusoid and therefore does not give a secular
drift. ${\vec u}_2$ is a sinusoid that grows in amplitude
and therefore does not give a secular drift either,
because if at any time we wait one period longer, the
displacement from the starting point given by ${\vec
u}_2$ is in the opposite direction and even larger; the
displacement given by ${\vec u}_2$ is not proportional to
$t$. If now we compute the trajectory of an element of
fluid by solving $d{\vec x}/dt = {\vec u}({\vec x},t)$,
we find that the only term that is proportional to
$\alpha^2t$ arises from the $\alpha {\vec u}_1$ term in
the Eulerian velocity field and from the fact that the
velocity field at the position of the fluid element
varies as it moves. The {\it new} part of the Eulerian
velocity field that arises from solving the fluid
equations to second order in $\alpha$ does not contribute
to the secular drift to ${\cal O}(\alpha^2)$; the secular
drift is already given {\it exactly} to this order by the
velocity field obtained by solving the {\it linearized}
fluid equations.

	We expect the drift of fluid elements that we
have computed from the $r$-wave velocity field
(\ref{eom}) to be qualitatively correct and perhaps exact
to ${\cal O}(\alpha^2)$.

\section{Magnetic field evolution equations: the Lagrangian approach}
\label{app_b}

	We here briefly show the derivation of the
integral form of the induction equation
(\ref{intgrl_indctn}) discussed in Section \ref{la}.
Consider a Lagrangian coordinate system ${\mathbf
x}$. The differential equation (\ref{induction_1}) can
then be written in terms of a generic vector field
${\mathbf A}$ as
\begin{equation}
\label{dxdt_1}
\frac{D }{D t} ( {\mathbf A}) = 
	({\mathbf A} \cdot \nabla) 
	\delta {\mathbf v} \ .
\end{equation}
Equation (\ref{dxdt_1}) can be readily integrated for an
infinitesimal time interval $\delta t$ from an initial
time $t_0$, so that
\begin{equation}
\label{dxdt_2}
A^i(t_0 + \delta t) =  A^i(t_0) + 
	\left [ 
	 A^j(t_0) \frac{\partial v^i}{\partial x^j} 
	\right ] \delta t = 
	 A^j(t_0) \frac{\partial }{\partial x^j} 
	\biggl [ x^i(t_0)+ v^i(t_0)\delta t \biggr ] =
	 A^j(t_0) \frac{\partial}{\partial x^j}
	\bigl[ {\tilde x}^i(t_0 + \delta t)\bigr] \ ,
\end{equation}
thereby relating the value of ${\mathbf A}$ at the new
time $t + \delta t$, to the value of ${\mathbf A}$ at
$t_0$. The extension of (\ref{dxdt_2}) to a finite time
interval is straightforward and yields
\begin{equation}
\label{intgrl_indctn_a}
A^j ({\tilde {\mathbf x}}, t) = 
	A^k ({\mathbf x}, t_0) 
	\left[\delta^j_{\ k} + 
	\frac{\partial \Delta {\tilde x}^j(t)}{\partial x^k(t_0)} 
	\right]	=
	A^k ({\mathbf x}, t_0) 
	\frac{\partial {\tilde x}^j(t)}{\partial x^k(t_0)} 
	\ ,
\end{equation}
where $\Delta {\tilde x}^j(t) = {\tilde x}^j(t) - {\tilde
x}^j(t_0) = {\tilde x}^j(t) - {x}^j(t_0)$.

\end{document}